# Comment on 'first accuracy evaluation of NIST-F2'

Kurt Gibble

Department of Physics, The Pennsylvania State University, University Park, PA 16802, USA

We discuss the treatment of the systematic frequency shifts due to microwave lensing and distributed cavity phase in "First accuracy evaluation of NIST-F2" 2014 Metrologia **51** 174–182. We explain that the microwave lensing frequency shift is generally non-zero and finite in the limit of no applied microwave field. This systematic error was incorrectly treated and we find that it contributes a significant frequency offset. Accounting for this shift implies that the measured microwave amplitude dependence (e.g due to microwave leakage) is comparable to the total reported inaccuracy. We also discuss the importance of vertically aligning the fountain perpendicular to the axis of the cavity feeds, when the cavity has only two independent feeds. Finally, we note that background gas collisions have a different behavior for cold clock atoms than for clock atoms at room-temperature, and therefore room temperature measurements do not directly apply to laser-cooled clocks.

We comment on several aspects of the recent accuracy evaluation of NIST-F2 [1], which reported a total systematic uncertainty of $1.1 \times 10^{-16}$. Our most significant remark regards the evaluation of the microwave lensing frequency shift [2-6], which was briefly described in [1] and more fully presented in a recent preprint [7]. They calculated a frequency offset of $0.2 \times 10^{-16}$ that goes to zero in the limit of zero microwave amplitude. We show that this frequency shift of NIST-F2 is larger, of order $0.9 \times 10^{-16}$, and comparable to other evaluations, which range from $0.6 \times 10^{-16}$ to $0.9 \times 10^{-16}$ [4-6]. In [1] Heavner et al. included microwave lensing with other systematic errors that depend on microwave amplitude, such as microwave leakage. When the shift that we calculate is removed from their measured amplitude dependence, the clock's frequency has a significant amplitude dependence with an offset at optimal amplitude from other amplitude dependent shifts (e.g. microwave leakage) that is comparable to the total systematic uncertainty reported in [1]. We discuss the amplitude dependence of the microwave lensing frequency shift and clarify points about frequency shifts from distributed cavity phase (DCP) [4-6,8-10], particularly $m=1$ phase gradients, and background gas collisions [11-15].

Heavner et al. considered that the amplitude dependence of "any microwave lensing shift scales just like the microwave leakage shift," which is proportional to $\delta\nu/\nu = A\, n\, \sin(n\pi/2) = A'\, \phi\, \sin(\phi)$ (dashed curve in Fig. 1a and, in [1], Fig 6), reporting measurements with Rabi pulse areas of $\phi = n\pi/2$ for $n=1,3,5,\ldots$ [1,7,14,16]. Previously, we have shown that the microwave lensing shift scales as $\delta\nu/\nu \propto \phi_1/\sin(\phi_1)$ [2-4], where $\phi_{1(2)}$ is the Rabi pulse area in the first (second) Ramsey interaction. A useful approximation, applied in [4-6], following (5) in [2], neglects the small transverse variation of $\sin(\phi_{1,2})$ near $\phi_{1,2} = \pi/2$ [3]:

$$\frac{\delta\nu}{\nu_R} = \frac{\phi_1}{\sin(\phi_1)} \frac{a(t_{2L}-t_1)}{k(t_2-t_1)} \cdot \frac{\int_0^{2\pi}\int_{r_{1L}<a} \frac{r_{2L0}(t_1-t_{1L}) + r_{1L}(t_{2L}-t_1)\cos(\phi_{2L0})}{r_1(t_{2L}-t_{1L})} J_1(kr_1) \psi_0^2(\vec{r}_{2L0},\vec{r}_{1L}) \Big|_{r_{2L0}=a} d\vec{r}_{1L} d\phi_{2L0}}{\int_{r_{2L0}<a}\int_{r_{1L}<a} \psi_0^2(\vec{r}_{2L0},\vec{r}_{1L}) d\vec{r}_{1L} d\vec{r}_{2L0}} \quad . \quad (1)$$

$$\psi_0^2(\vec{r}_{2L0},\vec{r}_{1L}) = e^{-\frac{r_{2L0}^2 w_{1L}^2 + r_{1L}^2 w_{2L}^2 - 2r_{2L0}r_{1L}(w_0^2 + u^2 t_{2L}t_{1L})\cos(\phi_{2L0})}{w_0^2 u^2 (t_{2L}-t_{1L})^2}} \qquad \vec{r}_1 = \frac{\vec{r}_{2L0}(t_1-t_{1L}) + \vec{r}_{1L}(t_{2L}-t_1)}{t_{2L}-t_{1L}}$$

Here, $a$ is the radius of the microwave cavity's aperture, $\cos(\phi_{2L0}) = \hat{r}_{2L0} \cdot \hat{r}_{1L}$, $t_1$ and $t_2$ are the times of the Ramsey interactions, $t_{1L}$ ($t_{2L}$) is the time that the atoms pass through the most restrictive aperture that is before (after) the Ramsey interactions, $u = (2k_BT/m)^{1/2}$ is the thermal velocity spread, $w_0$ is the initial $e^{-1}$ cloud radius, $w_{1L,2L}^2 = w_0^2 + u^2 t_{1L,2L}^2$, and $\nu_R = h\nu^2/2mc^2$ is the recoil shift of a microwave photon. The transverse variation of $\sin(\phi_{1,2})$ and detection inhomogeneities usually give small corrections for $\phi_{1,2}$ near $\pi/2$ and noticeable deviations at higher microwave amplitude (see Fig. 1b) [4-6].

Our results [2-6], including (1), yield a non-zero microwave lensing frequency shift in the limit of zero first pulse area $\phi_1 \to 0$, whereas [1] considers a frequency shift that scales as $\phi_1^2$ (for $\phi_1 = \phi_2$) and asserts that a non-zero shift for $\phi_1 \to 0$ is unphysical [1,7]. The

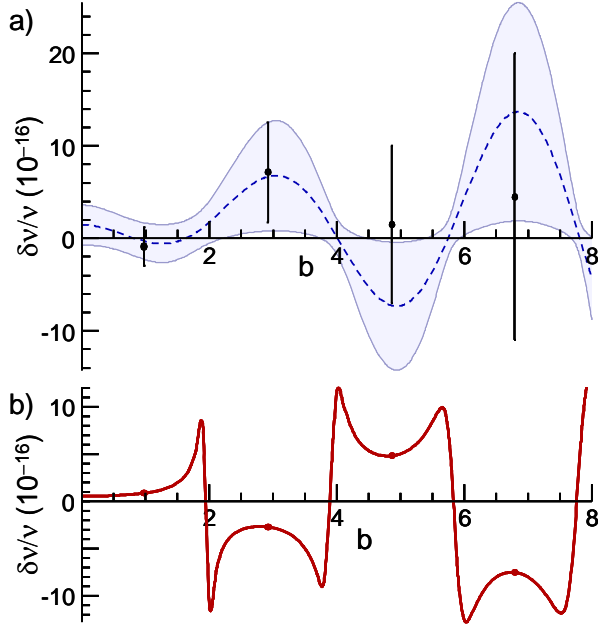

**Figure 1** a) Measured frequency of NIST-F2 versus microwave amplitude b for n=1,3,5,7 π/2 pulses [1], corrected for the calculated microwave lensing frequency shifts in b). The dashed curve and shaded region are a fit and uncertainty to A n sin(nπ/2) plus an offset with A = $-1.7\times10^{-16}\pm1.5\times10^{-16}$. b) Calculated microwave lensing frequency shift [2-6] of NIST-F2. The dots indicate 1,3,5,7 π/2 pulses.

origin of the microwave lensing shift is the impulse delivered to the atoms during the first Ramsey interaction by the gradient of the dipole energy of the atom in the microwave standing wave [2-4,7]. The perturbation of the transition probability $\delta P \propto \phi_1 \sin(\phi_2)$ [2-4] is thus linearly proportional to the deflection of the atomic population during the first Ramsey pulse and goes to 0 for $\phi_1 \rightarrow 0$. However, the Ramsey fringe amplitude $\Delta P_R$ is proportional to $\sin(\phi_1)\sin(\phi_2)$ and also goes to 0 for $\phi_1 \rightarrow 0$. Therefore, we get that the frequency shift $\delta\nu/\nu = \delta P/ \pi\Delta P_R (t_2-t_1)$ is generally non-zero and finite as $\phi_1 \rightarrow 0$. We thus believe this limit is the expected physical behavior. A related well-known frequency shift is a photon recoil shift of an optical transition [17]. In microwave clocks, the atomic wavefunctions are localized to less than a wavelength of the microwave standing wave. Integrations of the Schrödinger equation [2,18] apply to large wave packets that span many wavelengths, in which case the dipole forces from a standing wave produce resolved photon recoils. In this regime, the microwave lensing shift becomes a photon recoil shift [2,18]. In optical spectroscopy, photon recoil shifts are discreet and thus independent of field amplitude, remaining non-zero as the Rabi pulse area ϕ goes to 0, as in (1).

Equation (1) yields a microwave lensing shift of $0.87\times10^{-16}$ for typical and estimated parameters of NIST-F2[*]. Including the transverse variation of $\sin(\phi_{1,2})$ [4-6] yields the curve in Fig. 1b, and no significant difference from Eq. (1) at optimal amplitude. This microwave lensing shift is comparable to that of other fountains and significantly greater than $0.17\times10^{-16}$ calculated in [7]. Subtracting our predicted microwave lensing shift from the measurements reveals a significant amplitude dependence of the clock's frequency in Fig. 1a, potentially due to microwave leakage [1]. Fitting the data in Fig. 1a to $\delta\nu/\nu$= A n sin(n π/2) + $y_0$ yields a $\chi^2$ of 0.7 for A= $-1.7\times10^{-16}$ and $y_0$= $1.5\times10^{-16}$ and $\chi^2$ increases by 1 for $\delta A=\pm1.5\times10^{-16}$. These give an amplitude dependent shift (e.g. microwave leakage) at optimal power, and an associated uncertainty δA, that is greater than the assigned $0.8\times10^{-16}$ uncertainty for microwave amplitude dependent shifts and the reported total of all systematic uncertainties of $1.1\times10^{-16}$ [1]. If the microwave lensing shift is conservatively assigned an uncertainty of ±50% (or ±100%) [4-6], this changes A by $\pm0.3\times10^{-16}$ (or $\pm0.5\times10^{-16}$) when this range of calculated microwave lensing shifts is removed from the data, as in Fig. 1a. Even though the microwave lensing shift deviates noticeably from the n sin(n π/2) model for microwave leakage at n=1,3,5, and 7, $\chi^2$ remains sufficiently small, even for a ±100% uncertainty for microwave lensing. Thus, an uncorrelated theoretical uncertainty as large as $\pm0.87\times10^{-16}$ would slightly increase the uncertainty of the total microwave amplitude dependent shift (e.g. microwave leakage plus microwave lensing) at optimal amplitude to $\pm1.6\times10^{-16}$.

---

[*] From [1], we consider the following parameters: a cloud FWHM of 3 cm, a temperature of 0.46 μK, an interrogation time of 0.6s, 1 cm apertures that are 4.6 cm and 48 cm above the molasses region, a Ramsey cavity that is 63 cm above the molasses region, and $\phi_1$= 1.09 π/2 for a π/2 pulse. The fountain dimensions that we estimate, and their uncertainties, could not be confirmed; private communication, S. Jefferts (2014). However, NIST provided 'The "launch position" is 0.606 m below the lower aperture.' (NIST Response to Freedom of Information Act request 2014-001422). Dimensions in this range, with the Ramsey cavity between 63 and 78 cm above the molasses region, neither change the microwave lensing shifts we calculate by more than 5% nor the uncertainty of the fit in Fig. 1a.

The treatment by Ashby *et al.* [7] is a Schrödinger evolution, considering the same leading effects as in [2,3]. They also include some normally negligible contributions, such as small phase shifts during the Ramsey interactions and the momentum changes during the second Ramsey interaction [3,7]. It is important to try to address some differences between these calculated frequency shifts, especially because this systematic shift has not yet been observed experimentally and it is significant in comparison to the inaccuracies reported in [1,4-6,14]. We therefore note that Ashby *et al.* mistakenly treated the microwave field of their cavity, using the solution for a cavity with no holes in the endcaps, $H_z \propto J_0(\gamma \rho)\sin(k_z z)$ [7], where $\gamma \approx 3.83/R$ and R is the 30 mm cavity radius. This gives an incorrect position-dependent Rabi pulse area of $\phi_1(\rho) = \phi_1(0) J_0(\gamma \rho)$. An integration of Maxwell's equations, given in [9], used in [4-6], and explicitly reproduced here, generally treats the perturbation of the field by the holes. With a Rabi tipping angle

$$\phi_1(\vec{r}) = \int_{-\infty}^{\infty} H_z(\vec{r}, z)\, dz, \quad (2)$$

we integrate the wave equation $(\nabla^2 + k^2)H_z(\mathbf{r},z)=0$ over z, from a region with no field, through the cavity, to a region again with no field:

$$\begin{aligned}0 &= \int_{-\infty}^{\infty}\left(\nabla_{tr}^2 + \partial_z^2 + k^2\right)H_z(\vec{r},z)\,dz \\ &= \left(\nabla_{tr}^2 + k^2\right)\phi_1(\vec{r}) + \int_{-\infty}^{\infty}\partial_z^2 H_z(\vec{r},z)\,dz\end{aligned} \quad (3)$$

Here the integral in the last term is 0, giving a transverse wave equation for $\phi_1(\mathbf{r})$, with $k=\omega/c$ and $\omega=2\pi\times 9.192\ldots$ GHz. For an azimuthally symmetric $H_z(\rho,z)$, we get $\phi_1(\rho)=\phi_1(0) J_0(k \rho)$, independent of the cavity geometry [9]. Thus, $J_0(k \rho)$ has a 2.3 times larger curvature and gradients of the dipole energy from the microwave field than $J_0(\gamma \rho)$ for a cavity radius of 3 cm, and a similarly larger microwave lensing frequency shift. This accounts for some of the differences between [1,7] and [4-6]. On this same point, in the discussion of microwave leakage, [1] also incorrectly considered that their "state selection cavity has considerably less variation of the tipping angle versus radial position than smaller diameter cavities." From above, all cavities have the same variation of tipping angles when microwaves are applied for the entire cavity traversal. Moreover, smaller selection cavity radii allow pulses to be applied when the atoms are vertically centered in a cavity [4-6], giving a somewhat smaller spatial variation of the pulse area, very close to $J_0(\gamma \rho)$.

Regarding "the magnetic field defining equation $(B(\mathbf{r})=\cos(k_{1x}x)\cos(k_{1z}z))$ has no units and no amplitude factor" in [1], we note that [2] included a concise and general treatment of both electric and magnetic dipole interactions (see footnote [12] in [2]). The omitted constants that relate magnetic field and Rabi pulse area are well-known and given in several of the references of [1], including [4,8,9], where [4] explicitly gives the subsequent corresponding equations of [6].

Concerning distributed cavity phase (DCP) frequency shifts, linear (m=1) phase gradients have a naturally large scale [8-10]. Fountain clock cavities have opposing feeds so that phase gradients from the feeds can be eliminated by adjusting the amplitudes of the feeds [8-10]. Tilting the fountain probes the phase gradient along the tilt direction and the frequency difference between supplying power alternately to the two opposing feeds measures the difference of phase gradients produced by the feeds, independent of the phase gradients produced by losses in the cavity. The power dependence of the DCP tilt sensitivity can be used to vertically align the fountain along the axis of these opposing feeds [1,4-6,9,10,19]. Heavner *et al.* reported tilt measurements in two orthogonal directions and found very similar results for both directions. Because their fountain only has two independent feeds, the tilt measurements described in [1] can only probe the difference of the phase gradients of the two opposing sets of feeds. There must exist an axis that has no difference of linear phase gradients from the feeds and the power dependence of the phase gradients from the feeds cannot be used to vertically align the fountain along this axis. Thus the 100 μrad vertical uncertainty from the vertical alignment can only be applied along one of two orthogonal tilt directions and another method is needed to find vertical along the orthogonal direction [4-6,10,20]. As a result, the uncertainty from the direction perpendicular to the opposing feeds has been the dominant DCP uncertainty in previous evaluations [4-6,10].

With respect to the history of DCP shifts in [1], we disagree that [9,10] extended the earlier work of [21] on DCP shifts. Ref [8] solved for the three dimensional phase variations of cylindrical cavities and rigorously defined the effective phase of the cavity's field and treated an atom traversing the cavity. The significantly earlier measurements and calculations by Lemonde *et al.* [22] motivated the analysis in [8,9,23,24] of the power dependence of frequency shifts due to longitudinal phase gradients, using the phase variations calculated in [8]. In [8], it was written that the transverse variation of the tipping angle is essential to understand and correctly calculate the power dependence of the azimuthally symmetric (m=0) DCP shifts [8,9]. As discussed previously [9,25], the phenomenological field used in [21] was inconsistent with Maxwell's equations and the published solutions for cavities [8], and [21] did not include the required transverse variation of the tipping angle. The proper amplitude dependence of the azimuthally symmetric DCP shift was first reported in [23,24], and later more fully in [9]. More important than clarifying the history, we want to emphasize that the power dependence of the m=0 DCP shifts is no longer particularly helpful for the accuracy evaluations of fountain clocks [6,9,10]. For example, [1] did not report frequency measurements versus microwave amplitude that include the important amplitudes for the m=0 DCP shifts, near 4, 6, and 8 $\pi/2$ pulses. Calculations of the m=0 DCP shifts are sufficient for evaluations because the measurements versus amplitude in [10] have stringently tested the model of the azimuthally symmetric DCP shift (and also calculations of m≥1 phase gradients). While the m=0 DCP shift in [1] is undoubtedly negligible for $\phi_{1,2}=\pi/2$ pulses, [1] does not provide a description of an analysis.

Lastly, we note the frequency shift from background gas collisions for cold clock atoms has different physics than for room-temperature clock atoms, because cold atoms with large scattering phase shifts are ejected from the fountain [11-15]. As a result, the measured frequency shifts for room-temperature atoms used in [1] cannot readily be applied to a cold atom clock.

In summary, the microwave lensing frequency shift, m=1 phase gradients, and background gas shifts were not properly evaluated in [1]. The microwave lensing frequency shift that we calculate implies that the clock's frequency versus microwave amplitude has significant frequency shifts. At optimal amplitude these shifts and their associated uncertainty of $\pm1.5\times10^{-16}$ to $\pm1.6\times10^{-16}$ are comparable or larger than the total systematic uncertainty that was reported.

We thank S. Jefferts for providing the numerical data of clock frequency versus microwave amplitude and gratefully acknowledge financial support from NASA, the NSF, and Penn State.